\documentclass[aps,twocolumn,showpacs,preprintnumbers,amsmath,amssymb,superscriptaddress,floatfix,nofootinbib]{revtex4}
\usepackage{lipsum}
\usepackage{graphicx}
\usepackage{epsfig}
\usepackage{epstopdf}
\usepackage{hyperref}
\usepackage{amsmath}
\usepackage{amsfonts}
\usepackage{amssymb}

\begin{document}

\title{$\Lambda_b \to \pi^- (D_s^-) \Lambda_c(2595),~\pi^- (D_s^-) \Lambda_c(2625)$ decays and $DN,~D^*N$ molecular components}

\author{Wei-Hong Liang}
%\email{liangwh@gxnu.edu.cn}
\affiliation{Department of Physics, Guangxi Normal University,
Guilin 541004, China}
\affiliation{Departamento de
F\'{\i}sica Te\'orica and IFIC, Centro Mixto Universidad de
Valencia-CSIC Institutos de Investigaci\'on de Paterna, Aptdo.
22085, 46071 Valencia, Spain}

\author{Melahat Bayar}
%\email{xiejujun@impcas.ac.cn}
\affiliation{Department of Physics, Kocaeli University, 41380, Izmit, Turkey}

\author{Eulogio Oset}
%\email{oset@ific.uv.es}
 \affiliation{Departamento de
F\'{\i}sica Te\'orica and IFIC, Centro Mixto Universidad de
Valencia-CSIC Institutos de Investigaci\'on de Paterna, Aptdo.
22085, 46071 Valencia, Spain}

\date{\today}

\begin{abstract}
From the perspective that the $\Lambda_c(2595)$ and $\Lambda_c(2625)$ are dynamically generated resonances from the $DN,~D^*N$ interaction and coupled channels, we have evaluated the rates for $\Lambda_b \to \pi^- \Lambda_c(2595)$ and $\Lambda_b \to \pi^- \Lambda_c(2625)$ up to a global unknown factor that allows us to calculate the ratio of rates and compare with experiment, where good agreement is found. Similarly, we can also make predictions for the ratio of rates of the, yet unknown, decays of $\Lambda_b \to D_s^- \Lambda_c(2595)$ and $\Lambda_b \to D_s^- \Lambda_c(2625)$ and make estimates for their individual branching fractions.
\end{abstract}

\maketitle

\section{Introduction}

The weak decay of $B$ and $D$ mesons, as well as that of $\Lambda_b,\Lambda_c$ baryons, has brought an unexpected source of information on the nature of many hadrons which are produced in the final states (see recent reviews in Ref. \cite{osetreview}), adding new elements into the debate on the structure of hadrons \cite{klempt,crede}. The reactions that triggered these studies were the $B^0 \to J/\psi \pi^+ \pi^-$ and $B_s^0 \to J/\psi \pi^+ \pi^-$ observed in LHCb \cite{lhcb}. In the first reaction the $\pi^+ \pi^-$ gave rise to the $f_0(500)$ and there was only a very weak signal of the $f_0(980)$, while in the second reaction the $f_0(980)$ excitation was very pronounced and there was no signal of the $f_0(500)$. These results were soon interpreted within the context of the chiral unitary approach in Ref.~\cite{liangb}, where the $f_0(500)$ and $f_0(980)$ appear as a consequence of the pseudoscalar meson-pseudoscalar meson interaction in coupled channels \cite{npa}, using dynamics from the chiral Lagrangians \cite{gasser}. The same idea, with a different formalism, has been applied later with the same conclusions \cite{hanhart,wwang}.

$\Lambda_b$ decays followed in this line, and in Ref.~\cite{rocamai} the $\Lambda_b \to J/\psi \Lambda(1405)$ decay was studied, making predictions for $\pi \Sigma$ and $\bar K N$ invariant mass distributions. The predictions for the $s$-wave $K^- p$ mass distribution, associated to the $\Lambda(1405)$, were corroborated in the posterior experimental study of this reaction by the LHCb collaboration, in the experiment where two pentaquark signals were found \cite{qentaex}. Related work followed in Ref.~\cite{hyodo} in the weak decay of the $\Lambda_c$ into $\pi^+$ and a pair of meson-baryon states, $MB$, which gives rise to the $\Lambda(1405)$ and the $\Lambda(1670)$. Similarly, in Ref.~\cite{feijoo} the $\Lambda_b \to J/\psi K \Xi$ reaction was studied, which sheds light on the pseudoscalar-baryon interaction at energies above the $\Lambda(1405)$ region. More recently the $\Xi_c \to \pi^+ MB$ reaction has also been shown to be a good tool to investigate the $\Xi (1620)$ and $\Xi (1690)$ resonances \cite{miyahara}. Related reactions aimed at the production of pentaquark states have been reviewed in Refs.~\cite{chen,slzhu,pentarev}.

In the present work we study the $\Lambda_b \to \pi^- \Lambda_c(2595), ~\pi^- \Lambda_c(2625)$, $\Lambda_b \to D_s^- \Lambda_c(2595)$ and $\Lambda_b \to D_s^- \Lambda_c(2625)$ reactions and make predictions for the ratios of the branching fractions for the first two and last two reactions. Also, using the experimental values of the branching ratios for the first two decay modes, we make predictions for the branching fractions of the last two reactions. The starting point of our study is the assumption that the $\Lambda_c(2595)$ and $\Lambda_c(2625)$ states are dynamically generated from the pseudoscalar-baryon and vector-baryon interaction, and particularly from the $DN$ and $D^* N$ channels. The $\Lambda_c(2595)$ ($J^P=1/2^-$) has much resemblance to the $\Lambda(1405)$, and can be thought as being obtained by substituting the strange quark by a $c$ quark. The history of the $\Lambda(1405)$ is long (see review in the PDG \cite{tetsuoulf}). It appears dynamically generated from the interaction of $\bar K N,~\pi \Sigma$ and other coupled channels, and there are two states in the vicinity of the nominal mass \cite{ollerulf,cola}.

Within the picture of dynamically generated resonances, the $\Lambda_c(2595)$ was obtained in Ref.~\cite{hofmann} from the interaction of pseudoscalar-baryon channels, $DN$ and $\pi \Sigma_c$, essentially. The formalism was simplified and improved in Ref.~\cite{mizutani}. A step forward was given in Ref.~\cite{carmen}, were vector-baryon states, in particular $D^*N$ where added as coupled channels. An SU(8) spin-flavour symmetry scheme was used and the
$\Lambda_c(2595)$ was obtained with a large coupling to the $D^*N$. Further steps were given in Ref.~\cite{romanets}, where once again the SU(8) scheme was used, with some symmetry breaking to match an extension of the Weinberg-Tomozawa interaction in SU(3). Among other resonances, the $\Lambda_c(2595)$ ($J^P=1/2^-$) and the $\Lambda_c(2625)$ ($J^P=3/2^-$) were obtained.

Further work to include the vector-baryon states was done in Ref.~\cite{uchino}, where following the work of Ref.~\cite{garzon} in the light sector, a microscopic picture for $DN,~D^*N$ transition based on pion exchange was used. The state $\Lambda_c(2595)$ was obtained in $s$-wave, coupling both to $DN$ and $D^*N$, and the $\Lambda_c(2625)$, with $J^P=3/2^-$, was obtained from $D^*N$ and other coupled channels of vector-baryon type, with the largest coupling to $D^*N$.

In the present work we shall be able to show that both the $DN$ and $D^*N$ components are relevant in the $\Lambda_b \to \pi^- \Lambda_c(2595)$ and also we can relate this reaction to the $\Lambda_b \to \pi^- \Lambda_c(2625)$. We shall also see that the rates obtained are very sensitive to the relative sign of the coupling of this resonance to $DN$ and $D^*N$, and how the proper sign gives rise to results compatible with experiments. In addition, the formalism developed here allows one to obtain the branching ratios for $\Lambda_b \to D_s^- \Lambda_c(2595)$ and $\Lambda_b \to D_s^- \Lambda_c(2625)$ from those of $\Lambda_b \to \pi^- \Lambda_c(2595)$ and $\Lambda_b \to \pi^- \Lambda_c(2625)$ respectively.

\section{Formalism}

\begin{figure}[tb]\centering
\includegraphics[scale=0.56]{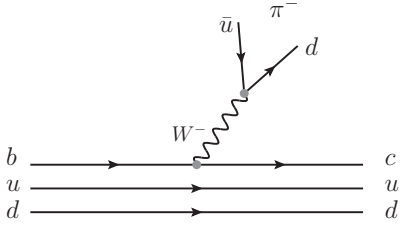}
\caption{Basic diagram for $\Lambda_b \to \pi^- \Lambda_c(2595)$. The $u$ and $d$ quarks are spectators and in isospin and strangeness $I=0$,~$S=0$.
\label{fig:FeynmanDiag1}}
\end{figure}

The basic diagram for the $\Lambda_b \to \pi^- \Lambda_c(2595)$ decay is shown in Fig. \ref{fig:FeynmanDiag1}.

\begin{figure}[tb]\centering
\includegraphics[scale=0.56]{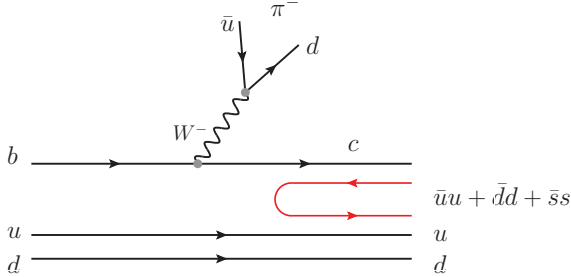}
\caption{Hadronization creating the $\bar qq$ pairs.
\label{fig:FeynmanDiag2}}
\end{figure}

The weak transition occurs on the $b$ quark, which turns into a $c$ quark, and a $\pi^-$ is produced through the mechanism of external emission \cite{chau}. Since we will have a $1/2^-$ or $3/2^-$ state at the end, and the $u,~d$ quarks are spectators, the final $c$ quark must carry negative parity and hence must be in an $L=1$ level. Since the $\Lambda_c(2595)$ and $\Lambda_c(2625)$ come from meson-baryon interaction in our picture, we must hadronize the final state including a $\bar q q$ pair with the quantum numbers of the vacuum. This is done following the work of Ref.~\cite{rocamai}. We include the $\bar uu+\bar dd+\bar ss$ as in Fig. \ref{fig:FeynmanDiag2}.
The $c$ quark must be involved in the hadronization, because it is originally in an $L=1$ state, but after the hadronization produces the $DN$ state, the $c$ quark in the $D$ meson is in an $L=0$ state.

The original state is
\begin{equation}\label{eq:Lambda_bState}
  | \Lambda_b \rangle =\frac{1}{\sqrt{2}}  | b(ud-du) \rangle,
\end{equation}
and after the weak process it becomes
\begin{equation}\label{eq:H}
 |H \rangle =\frac{1}{\sqrt{2}}  | c(ud-du) \rangle.
\end{equation}
The hadronization converts this state into $|H'\rangle$,
\begin{equation}\label{eq:Hprime}
 |H' \rangle =\frac{1}{\sqrt{2}}  | c(\bar uu+\bar dd+\bar ss)(ud-du) \rangle,
\end{equation}
which can be written as
\begin{equation}\label{eq:Hprime2}
 |H' \rangle =\frac{1}{\sqrt{2}} \sum_{i=1}^3 | P_{4i} \ q_i (ud-du) \rangle,
\end{equation}
where $P_{4i}$ is the $4i$ matrix element of the $q\bar q$ matrix in SU(4),
\begin{equation}\label{eq:P_matrix}
P\equiv(q \bar q)=\left(
           \begin{array}{cccc}
             u\bar u & u \bar d & u\bar s & u\bar c\\
             d\bar u & d\bar d & d\bar s & d\bar c\\
             s\bar u & s\bar d & s\bar s & s\bar c\\
             c\bar u & c\bar d & c\bar s & c\bar c\\
           \end{array}
         \right).
\end{equation}
The matrix can be written in terms of the physical mesons, pseudoscalar at the moment, and given in Ref.~\cite{danizou}
\begin{widetext}
\begin{equation}\label{eq:phimatrix}
M \to \phi \equiv \left(
           \begin{array}{cccc}
             \frac{1}{\sqrt{2}}\pi^0 + \frac{1}{\sqrt{3}}\eta + \frac{1}{\sqrt{6}}\eta' & \pi^+ & K^+ & \bar D^0\\
             \pi^- & -\frac{1}{\sqrt{2}}\pi^0 + \frac{1}{\sqrt{3}}\eta + \frac{1}{\sqrt{6}}\eta' & K^0 & D^-\\
            K^- & \bar{K}^0 & -\frac{1}{\sqrt{3}}\eta + \sqrt{\frac{2}{3}}\eta' &D_s^- \\
            D^0 & D^+ & D_s^+ &\eta_c \\
           \end{array}
         \right).
\end{equation}
\end{widetext}
Then Eq.  (\ref{eq:Hprime2}) can be written as
\iffalse
\begin{eqnarray}
   |H' \rangle &=& \frac{1}{\sqrt{2}} [ D^0 u(ud-du) + D^+ d(ud-du) \nonumber \\
   && + D^+_s s(ud-du) ].
\end{eqnarray}
\fi
\begin{equation}\label{eq:Hprime3}
 |H' \rangle =\frac{1}{\sqrt{2}} [ D^0 u(ud-du) + D^+ d(ud-du) + D^+_s s(ud-du) ].
\end{equation}
We can see that we have the three quarks in a mixed antisymmetric representation. Recalling that \cite{close}
\begin{align}\label{eq:LightBaryon}
&|p \rangle = \frac{1}{\sqrt{2}} |u(ud-du)\rangle, \nonumber \\
&|n\rangle = \frac{1}{\sqrt{2}} |d(ud-du)\rangle, \nonumber \\
&|\Lambda\rangle = \frac{1}{\sqrt{12}} |(usd-dsu)+(dus-uds)+2(sud-sdu)\rangle. \nonumber
\end{align}
We finally see that the hadronization has given rise to
\begin{eqnarray}\label{eq:Hprime4}
   |H' \rangle &=& |D^0 p +D^+ n +\sqrt{\frac{2}{3}} D^+_s \Lambda \rangle \nonumber \\
   &\simeq & \sqrt{2} |DN, I=0\rangle,
\end{eqnarray}
where we neglect the $D^+_s \Lambda$ that has a much higher mass than the $DN$ and does not play a role in the generation of the $\Lambda_c(2595)$. The isospin $I=0$ in Eq. (\ref{eq:Hprime4}) comes from the implicit phase convention in our approach, with the doublets $(D^+,\ -D^0)$ and $(\bar D^0,\ D^-)$.

\begin{figure}[tb]\centering
\includegraphics[scale=0.55]{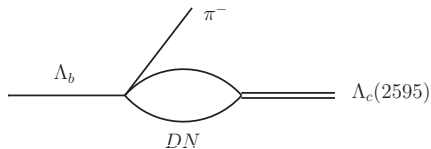}
\caption{Diagram to produce the $\Lambda_c(2595)$ through an intermediate propagation of the $DN$ state.
\label{fig:FeynmanDiag3}}
\end{figure}

The production of the resonance is done after the produced $DN$ in the first step merges into the resonance, as shown in Fig. \ref{fig:FeynmanDiag3}.
The transition matrix for the mechanism of  Fig. \ref{fig:FeynmanDiag3} gives us
\begin{equation}\label{eq:t_R}
  t_R=V_P \; \sqrt{2}\; G_{DN}\cdot g_{R,DN},
\end{equation}
where $V_P$ is a factor that includes the dynamics of $\Lambda_b \to \pi^- DN$, $G_{DN}$ is the loop function for the $DN$ propagation \cite{uchino}, and
$g_{R, DN}$ is the coupling of the resonance to the $DN$ channel in $I=0$ \cite{uchino}.

The width for the decay process is given by
\begin{equation}\label{eq:Gamma_R}
   \Gamma_R=\frac{1}{2\pi}\frac{M_{\Lambda_c^*}}{M_{\Lambda_b}} \overline{\sum}\sum\left| t_R \right|^2 p_{\pi^-},
\end{equation}
where $\overline{\sum}\sum$ stands for the sum and average over polarizations.

The arguments used above can be equally used for the production of $D^*N$. The $V_p$ factor would now be different, but in the next section we shall show how to relate them.

\section{Angular and spin matrix elements}
The discussion in the former section has only payed attention to the flavour aspect of the hadronization.
If we wish to relate the $DN$ and $D^* N$ production, we need to go in more detail into the problem and take into account explicitly the matrix elements involved. The first step is to consider the spin and angular dependence of the created pair. We want it in $J=0$, positive parity and positive $C$ parity. Since the parity of the antiquark is negative, we need it in $L=1$, which also forces the spin of the pair to be $S=1$, leading to the $^3P_0$ configuration \cite{micu,leyuanc,close}.

Since the $\bar q q$ pair has $J=0$ and so has the $ud$ spectator pair, the total angular momentum of the final meson-baryon state is given by the combination of the angular momentum and spin of the $c$ quark, and we have
\begin{equation}\label{eq:JM}
 |JM\rangle=\sum_m \mathcal{C}(1\frac{1}{2}J;\ m, M-m) Y_{1m}\left| \frac{1}{2}, M-m \right\rangle,
\end{equation}
where $\mathcal{C}(J_1 J_2 J;\ m_1, m_2, M)$ [or writing equivalently as $\mathcal{C}(J_1 J_2 J;\ m_1, M-m_1)$] is the
Clebsh-Gordan coefficient (CGC) combining $| J_1 m_1 \rangle$ and $| J_2 m_2 \rangle$ to get the $| J M \rangle$ state, and $Y_{lm}$ is the spherical harmonic.
On the other hand, the spin state of the $\bar qq$ pair is given by
\begin{equation}\label{eq:1S3}
 |1S_3\rangle=\sum_s \mathcal{C}(\frac{1}{2}\frac{1}{2}1;\ s, S_3-s) \left| \frac{1}{2},s \right\rangle \left| \frac{1}{2},S_3-s \right\rangle.
\end{equation}
We are only concerned about the angular momentum counting and can consider a zero range interaction, as done in a similar problem where the angular momentum is at stake, the pairing in nuclei \cite{brown,bhaduri}. Then we associate to the antiquark an angular momentum
$|1,M_3\rangle\equiv Y_{1M_3}$, and thus the $J=0$ $\bar q q$ wave function is given by
\begin{equation}\label{eq:00}
 |00\rangle=\sum_{M_3,S_3} \mathcal{C}(110;\ M_3, S_3,0) Y_{1M_3}| 1\ S_3\rangle,
\end{equation}
which requires $M_3+S_3=0$, $S_3=-M_3$, hence,
\begin{eqnarray}\label{eq:00new}
   |00\rangle&=&\sum_{M_3} \mathcal{C}(110;\ M_3, -M_3) \ Y_{1M_3}  \nonumber \\
   & \times&  \sum_s \mathcal{C}(\frac{1}{2} \frac{1}{2} 1;\ s, -M_3-s) | \frac{1}{2}, s\rangle | \frac{1}{2}, -M_3-s\rangle.~~~~~~
\end{eqnarray}
The final meson-baryon state is $|JM\rangle |00\rangle$, given by Eqs. (\ref{eq:JM}), (\ref{eq:00new}).
We can combine the two spherical harmonics (we use formulas of Ref. \cite{rose} in what follows)
\begin{eqnarray}\label{eq:harmonics}
   Y_{1m}\ Y_{1M_3}&=&\sum_{l} 3\left[ \frac{1}{4\pi (2l+1)} \right]^{1/2}   \nonumber \\
   & \times&  \mathcal{C}(11l;\ m, M_3) \ \mathcal{C}(11l;\ 000) \ Y_{l,m+M_3},~~~
\end{eqnarray}
where for parity reasons, only $l=0,2$ contribute, but we are only concerned about $l=0$, which is suited for pseudoscalar-baryon final states with the $s$-wave that we only consider, and all quarks in the ground state. Then
\begin{equation}\label{eq:YY}
  Y_{1m}Y_{1M_3} \rightarrow (-1)^m \frac{1}{4\pi} \delta_{M_3,-m}.
\end{equation}
Then $|JM\rangle\ |00\rangle$ is given, rearranging the CGC, by
%\begin{eqnarray}\label{eq:JM00}
%  |JM>|00>&=& -\frac{1}{\sqrt{3}} \frac{1}{4\pi} \sum_m \sum_s  \mathcal{C}(1\frac{1}{2}J;\ m, M-m)\nonumber \\
%  & \times &  \mathcal{C}(\frac{1}{2} \frac{1}{2} l; \ s, m-s) \ |\frac{1}{2}, M-m>\ |\frac{1}{2}, s>\ |\frac{1}{2}, m-s>.
%\end{eqnarray}
\begin{widetext}
\begin{equation}\label{eq:JM00}
  |JM\rangle|00\rangle= -\frac{1}{\sqrt{3}} \frac{1}{4\pi} \sum_m \sum_s  \mathcal{C}(1\frac{1}{2}J;\ m, M-m)
  \ \mathcal{C}(\frac{1}{2} \frac{1}{2} 1; \ s, m-s) \ |\frac{1}{2}, M-m\rangle\ |\frac{1}{2}, s\rangle\ |\frac{1}{2}, m-s \rangle.
\end{equation}
\end{widetext}
Finally we combine the spin states of the $c$ quark and the antiparticle as
\begin{equation}\label{eq:combine}
  |\frac{1}{2}, M-m\rangle|\frac{1}{2},s\rangle=\sum_j \mathcal{C}(\frac{1}{2} \frac{1}{2} j; \ M-m,s) |j, M-m+s\rangle
\end{equation}
such that $j$ will be the spin of the pseudoscalar $D$ meson ($j=0$) or the vector $D^*$ meson ($j=1$).
Since the $ud$ quarks have $s=0$, the state $|\frac{1}{2},m-s\rangle$ gives the spin of the baryon and we can write
\begin{eqnarray}\label{eq:jBaryon}
   &&|j,M-m+s\rangle\ |\frac{1}{2},m-s\rangle  \nonumber \\
   &=& \sum_{J'} \mathcal{C}(j\frac{1}{2}J';\ M-m+s, m-s)\ |J',M\rangle,
\end{eqnarray}
where now $J'$ will be the final angular momentum of the $DN$ system. Obviously $J'$ should be equal to $J$,
but this requires a bit of Racah algebra to show up.
The $|JM\rangle\ |00\rangle$ state can now be written as
\begin{eqnarray}\label{eq:JM00}
 &&|JM\rangle|00\rangle  \nonumber \\
 &=& -\frac{1}{\sqrt{3}} \frac{1}{4\pi} \sum_{m,s}  \mathcal{C}(1\frac{1}{2}J;\ m, M-m) \mathcal{C}(\frac{1}{2} \frac{1}{2} 1; \; s, m-s) \nonumber \\
  & \times &   \sum_{j,J'}  \mathcal{C}(\frac{1}{2} \frac{1}{2} j; \ M-m, s)
  \mathcal{C}(j \frac{1}{2} J'; \ M-m+s, m-s) \nonumber \\
 & \times &   |J',M\rangle.
\end{eqnarray}

Recombining the CGC and using their symmetry properties, we can use Eq. (6.5a) of Ref.~\cite{rose} and find
\begin{widetext}
\begin{equation}\label{eq:CombCGC}
\sum_s \mathcal{C}(1\frac{1}{2} \frac{1}{2}; \ m, -s) \ \mathcal{C}(\frac{1}{2} jJ'; \ m-s, M-m+s)
\ \mathcal{C}(\frac{1}{2} j \frac{1}{2}; \ -s, M-m+s)
=R_{\frac{1}{2} \frac{1}{2}} \ \mathcal{C}(1 \frac{1}{2} J'; \ m, M-m).
\end{equation}
\end{widetext}
where
\begin{equation}\label{eq:RCoef}
  R_{\frac{1}{2} \frac{1}{2}}\equiv 2 W(1\frac{1}{2} Jj;\frac{1}{2}\frac{1}{2} )
\end{equation}
in terms of the $W$ Racah coefficients.
The other sum over $m$ gives now $J'=J$
\begin{equation}
  \sum_m \mathcal{C}(1\frac{1}{2} J; \ m, M-m) \ \mathcal{C}(1\frac{1}{2} J'; \ m, M-m) =\delta_{JJ'}
\end{equation}
such that finally
\begin{eqnarray}\label{eq:JM00New}
 &&|JM\rangle|00\rangle  \nonumber \\
 &=& \frac{1}{4\pi} \sum_{j} (-1)^{j-J+1/2} \sqrt{2j+1} W(1\frac{1}{2} Jj;\frac{1}{2}\frac{1}{2} )  \nonumber \\
 &~~~~~~& \times  |JM, \text{meson-baryon}\rangle \nonumber \\
 &\equiv& \sum_j  \mathcal{C}(j,J) |JM, \text{meson-baryon}\rangle.
\end{eqnarray}

Evaluating the Racah coefficients with formulas of the Appendix of Ref.~\cite{rose}, we have the results shown in table \ref{tab:CjJ}.
\begin{table}[tbH]
     \renewcommand{\arraystretch}{1.5}
     \setlength{\tabcolsep}{0.4cm}
\centering
\begin{tabular}{r|cc}
$\mathcal{C}(j,J)$ & $J=1/2$  &  $J=3/2$   \\
\hline
(pseudoscalar)~$j=0$  & $ \frac{1}{4\pi} \frac{1}{2}$  & 0   \\
(vector)~$j=1$  &$ \frac{1}{4\pi} \frac{1}{2\sqrt{3}}$  & $- \frac{1}{4\pi} \frac{1}{\sqrt{3}}$
\end{tabular}
\caption{$\mathcal{C}(j,J)$ coefficients in Eq.~(\ref{eq:JM00New}).}
\label{tab:CjJ}
\end{table}

\section{Evaluation of the weak matrix elements}
Up to global factors which are the same for vector-baryon or pseudoscalar-baryon production, the relevant elements that we need are that the $W^- \to \pi^-$ production is of the type \cite{gasser2,scherer}
\begin{equation}\label{eq:L_wpi}
  \mathcal{L}_{W,\pi} \sim W^{\mu} \partial_{\mu} \phi,
\end{equation}
and the $bWc$ vertex of the type \cite{chau,osetreview}
\begin{equation}\label{eq:L_bwc}
  \mathcal{L}_{\bar qWq}=\bar q_{\rm fin} W_{\mu} \gamma^{\mu} (1-\gamma_5) q_{\rm in}.
\end{equation}
For small energies of the quarks, the relevant matrix elements are the $\gamma^0$ and the $\gamma^i \gamma_5 (i=1,2,3)$.
Combining Eqs.~(\ref{eq:L_wpi}) and (\ref{eq:L_bwc}), and using the nonrelativistic reduction of $\gamma^0, \gamma^i \gamma_5$, the weak external pseudoscalar meson production has the structure
\begin{equation}\label{eq:Vp}
  V_P \sim q^0 + \vec{\sigma} \cdot \vec{q},
\end{equation}
with $q^0, \vec{q}$ the energy and momentum of the pion and $\vec{\sigma}$ the Pauli spin matrix acting on the quarks.
Assume $\varphi_{\rm in} (r)$ is the $b$ quark  radial wave function %(including the $Y_{00}$ spherical harmonic for economy in the rotation)
and
$\varphi_{\rm fin} (r)$ the radial wave function of the $c$ quark, and take the state $|JM'\rangle$ of Eq.~(\ref{eq:JM}) for the $c$ quark. The space matrix element is given by
\begin{equation}\label{eq:rme}
 \int d^3 r \ \varphi_{\rm in} (r) \ \varphi^*_{\rm fin} (r) \ Y^*_{1m} (\hat r) \ e^{-i \vec{q} \cdot \vec{r}}\ Y_{00},
\end{equation}
where $e^{-i \vec{q} \cdot \vec{r}}$ stands for the plane wave function for the outcoming pion. By using the expansion of $e^{-i \vec{q} \cdot \vec{r}}$
\begin{equation}
  e^{-i \vec{q} \cdot \vec{r}} = 4\pi \sum_{l'} (-1)^{l'} j_{l'} (qr) \sum_{\mu} (-1)^{\mu} Y^*_{l'\mu}(\hat r) Y^*_{l',-\mu}(\hat q), \nonumber
\end{equation}
Eq.~(\ref{eq:rme}) gives
\begin{equation}\label{rme2}
  -\sqrt{4\pi}\ i \ Y^*_{1m} (q) \ ME(q),
\end{equation}
where
\begin{equation}\label{eq:ME}
  ME(q)\equiv  \int r^2 dr \ j_{1} (qr)\ \varphi_{\rm in} (r) \ \varphi^*_{\rm fin} (r).
\end{equation}
One could evaluate this $ME$ with some quark model, but given the fact that we only want to evaluate ratios of rates, that the momenta $q$ involved in the different transitions are very similar and that $\varphi_{\rm fin} (r)$ is the same for all of them, we shall assume $ME(q)$ to be the same for all these transitions.
Hence, the weak matrix element for the $q^0$ term of Eq.~(\ref{eq:Vp}) is (note that $Y^*_{1m}$ becomes $Y^*_{1,M'-M}$)
\begin{eqnarray} \label{eq:Vp_q0}
 &&\langle JM'|q^0|\frac{1}{2} M\rangle \nonumber \\
 &=& -\sqrt{4\pi} \ i q^0 \sum_m  \mathcal{C}(1\frac{1}{2} J; \ m, M'-m)  \nonumber \\
 &&\times \langle \frac{1}{2},M'-m|\frac{1}{2} M\rangle Y^*_{1,M'-M} (\hat q) \; ME(q) \nonumber \\
  &=& -\sqrt{4\pi} \;i q^0 \mathcal{C}(1\frac{1}{2} J; \ M'-m, M) Y^*_{1,M'-M} (\hat q) \;ME(q).\nonumber
\end{eqnarray}

In the case of $J=1/2$, it is practical to write this matrix element in terms of the macroscopical $\vec \sigma \cdot \vec{q}$ operators, where
$\vec \sigma$ is acting not within quarks but within the baryon states $\Lambda_b$ and $\Lambda^*_c$.
Using the  Wigner-Eckart theorem and $\vec \sigma \cdot \vec{q}=\sum_\mu (-1)^\mu \sigma_\mu q_{-\mu}$,
with $\mu$ indices in spherical basis, $q_{-\mu}=q \sqrt{\frac{4\pi}{3}}\;Y_{1,-\mu}(\hat q)$, we have
\begin{eqnarray} \label{eq:Vp_sigq}
 &&\langle \frac{1}{2} M'|\vec \sigma \cdot \vec{q}|\frac{1}{2} M\rangle \nonumber \\
 &=& -\sqrt{4\pi} \ q \ \mathcal{C}(1\frac{1}{2} \frac{1}{2}; \ M'-M, M) Y^*_{1,M'-m}(\hat q).
\end{eqnarray}
We have ($q^0 = w_{\pi}$)
\begin{eqnarray}\label{eq:quarkLevel_macroLevel}
  &&J=1/2:  \nonumber \\
  &&q^0\mid_{\rm quark ~level} \to i\frac{w_{\pi}}{q} ME(q) \vec \sigma \cdot \vec{q} \mid_{\rm macroscopical ~level}.~~~~~~
\end{eqnarray}

In the case of $J=3/2$, we proceed in a similar way and introduce the macroscopical spin transition operator $\vec S^+$ from spin $1/2$ to $3/2$,
defined as
\begin{equation}\label{eq:SOperator}
  \langle \frac{3}{2}M'|\vec S^+ \cdot \vec q| \frac{1}{2} M\rangle =\mathcal{C}(\frac{1}{2}1\frac{3}{2}; \ M_\mu M'),
\end{equation}
which, via the Wigner-Eckart theorem implies a normalization of $S^+$ such that $\langle \frac{3}{2}||S^+||\frac{1}{2}\rangle\equiv 1$.
With this normalization we have the sum rule in Cartesian coordinates \cite{otw}
\begin{equation}\label{eq:SOperator2}
  \sum_{M'} S_i |M'\rangle \langle M'| S^+_j = \frac{2}{3}\delta_{ij} -\frac{i}{3}\epsilon_{ijk} \sigma_k.
\end{equation}
Then Eq.~(\ref{eq:Vp_q0}) can be cast at the macroscopical level through the substitution
\begin{eqnarray}\label{eq:quarkLevel_macroLevel}
  &&J=3/2: \nonumber \\
  &&\left. q^0\right|_{\rm quark ~level} \to -i\frac{w_{\pi}}{q} ME(q) \sqrt{3} \vec S^+ \cdot \vec q\mid_{\rm macroscopical ~level}.~~~~~
\end{eqnarray}

We must work now with the matrix element for the $\vec \sigma \cdot \vec q$ operator of Eq.~(\ref{eq:Vp}) at quark level.
By analogy to Eq.~(\ref{eq:Vp_q0}), the matrix element is now
\begin{eqnarray} \label{eq:Vp_sigmaq}
 &&\langle JM'|\vec \sigma \cdot \vec q|\frac{1}{2} M\rangle \nonumber \\
 &=& -\sqrt{4\pi} i  \sum_m  \mathcal{C}(1\frac{1}{2} J; \ m, M'-m)  \nonumber \\
 &&\times \langle \frac{1}{2},M'-m \left|\vec \sigma \cdot \vec q \right|\frac{1}{2} M\rangle Y^*_{1m} (\hat q) ME(q) \nonumber \\
  &=& -\sqrt{4\pi} i \sum_m \mathcal{C}(1\frac{1}{2} J; \ m, M'-m) (-) \sqrt{4\pi}q Y^*_{1m}(\hat q)\nonumber \\
  &&\times \mathcal{C}(1\frac{1}{2} \frac{1}{2}; \ M'-m-M, M)
  Y^*_{1,M'-m-M}(\hat q) ME(q),~~~~~
\end{eqnarray}
where in the last step we have used Eq.~(\ref{eq:Vp_sigq}).

Next we combine
\begin{widetext}
\begin{equation}%\label{eq:Vp_sigmaq3}
Y^*_{lm}(\hat q) Y^*_{l,M'-m-M} (\hat q) = \sum_{l'} 3 \left[ \frac{1}{4\pi (2l'+1)} \right]^{1/2}
\mathcal{C}(11l'; \ M'-m-M,m) \ \mathcal{C}(11l'; 000)
Y^*_{l', M'-M}(\hat q),
\end{equation}
\end{widetext}
which again only contribution for $l'=1,2$. We keep just the term with lowest angular momentum that should give the largest contribution, hence,
\begin{equation*}
  Y^*_{lm}(\hat q) Y^*_{l,M'-m-M} (\hat q) \to (-1)^m \frac{1}{4\pi} \delta_{MM'}.
\end{equation*}
Then Eq.~(\ref{eq:Vp_sigmaq}) becomes
\begin{eqnarray} \label{eq:Vp_sigmaq2}
 &&\langle JM'|\vec \sigma \cdot \vec q|\frac{1}{2} M\rangle \nonumber \\
 &\to& i \delta_{MM'} \sum_m (-1)^m \ \mathcal{C}(1\frac{1}{2} J; \ m, M-m)  q \nonumber \\
 &&\times \ \mathcal{C}(1\frac{1}{2} \frac{1}{2}; \ -m, M)  \nonumber \\
  &=& i   q \sum_m \mathcal{C}(1\frac{1}{2} J; \ m, M-m) \ \mathcal{C}(1\frac{1}{2} \frac{1}{2}; \ m, M-m) \nonumber \\
  &=& i  q \delta_{J,\frac{1}{2}},
\end{eqnarray}
where in the second last step we have permuted the last two angular momenta in $\mathcal{C}(1\frac{1}{2} \frac{1}{2}; \ -m, M)$ and changed the sign of the third components, which introduces the phase $(-1)^m$ that cancels the original $(-1)^m$ phase. We thus see that this term only contributes to $J=1/2$.

The study done allows us to write the weak vertex transition in terms of the following operator at the macroscopic level of the $\Lambda_b$ and $\Lambda^*_c$ baryons as
%\begin{widetext}
%\begin{equation}\label{eq:Operators}
%\left( i  \sqrt{4\pi} q + i\frac{w_\pi}{q}\sqrt{4\pi} \vec \sigma \cdot \vec q  \right) \delta_{J, 1/2} +\left(  i\frac{w_\pi}{q}\sqrt{12\pi} \vec S^+ \cdot \vec q   \right) \delta_{J, 3/2}.
%\end{equation}
%\end{widetext}
\begin{equation}\label{eq:Operators}
  \left( i   q + i\frac{w_\pi}{q} \vec \sigma \cdot \vec q  \right) \delta_{J, \frac{1}{2}} +\left( - i\frac{w_\pi}{q}\sqrt{3} \vec S^+ \cdot \vec q   \right) \delta_{J, \frac{3}{2}},
\end{equation}
where we have removed the factor $ME(q)$ in both terms.
If we combine this operator with the meson-baryon decomposition of Eq.~(\ref{eq:Vp_sigmaq}), we finally have a full transition $t$ matrix given, up to an arbitrary common factor, by
\begin{widetext}
\begin{equation}\label{eq:t_R2}
t_R= \left(  iq+ i \frac{w_\pi}{q} \vec \sigma \cdot \vec q \right) \left( \frac{1}{2} G_{DN}\ g_{R,DN} +\frac{1}{2\sqrt{3}} G_{D^*N}\ g_{R,D^*N}\right) \delta_{J, \frac{1}{2}}
+\left(  -i  \frac{w_\pi}{q} \sqrt{3}  \vec S^+ \cdot \vec q \right)\frac{1}{\sqrt{3}} G_{D^*N} \ g_{R,D^*N} \ \delta_{J, \frac{3}{2}}.
\end{equation}
\end{widetext}

Using Eq.~(\ref{eq:SOperator2}) and properties of the $\vec \sigma$ matrix,
it is easy to write now $\overline{\sum}\sum |t_R|^2$ in Eq.~(\ref{eq:Gamma_R}) as
\begin{widetext}
\begin{equation}\label{eq:t_R3}
\left[ \overline{\sum}\sum |t_R|^2 \right]_1= (q^2 + w^2_{\pi}) \left|  \frac{1}{2} G_{DN}\ g_{R,DN} +\frac{1}{2\sqrt{3}} G_{D^*N}\ g_{R,D^*N} \right|^2,~ {\rm for} ~J=\frac{1}{2};
\end{equation}
and
\begin{equation}\label{eq:t_R4}
\left[ \overline{\sum}\sum |t_R|^2 \right]_2 = 2 w^2_{\pi} \left|  \frac{1}{\sqrt{3}} G_{D^*N} \ g_{R,D^*N} \right|^2,~ {\rm for}~ J=\frac{3}{2}.
\end{equation}
\end{widetext}
\iffalse
%\begin{widetext}
\begin{equation}\label{eq:phimatrix}
\overline{\sum}\sum |t_R|^2 = \left\{
           \begin{array}{ll}
             (q^2 + w^2_{\pi}) \left|  \frac{1}{2} G_{DN}\ g_{R,DN} +\frac{1}{2\sqrt{3}} G_{D^*N}\ g_{R,D^*N} \right|^2, & {\rm for} J=\frac{1}{2};\\
             2 w^2_{\pi} \left|  \frac{1}{\sqrt{3}} G_{D^*N} \ g_{R,D^*N} \right|^2, & {\rm for} J=\frac{3}{2}.\\
           \end{array}
         \right.
\end{equation}
\end{widetext}
\fi

Eq.~(\ref{eq:Gamma_R}) gives then the partial decay widths up to an arbitrary normalization factor, the same for all the processes.

\section{Results}
The momentum of the pion in the decay is given by
\begin{equation}\label{eq:ppi}
  p_{\pi}= \frac{\lambda^{1/2}(M^2_{\Lambda_b}, m^2_\pi, M^2_R)}{2M_{\Lambda_b}}.
\end{equation}
On the other hand, the product $G_{DN}\cdot g_{R,DN}$ and $G_{D^*N} \cdot g_{R,D^*N}$ are tabulated in Ref.~\cite{uchino}. We copy these results in Table \ref{tab:Gcoupling}.
\begin{table}[tbH]
     \renewcommand{\arraystretch}{1.5}
     \setlength{\tabcolsep}{0.4cm}
\centering
\begin{tabular}{r|cc}
& $G_{DN}\cdot g_{R,DN}$  &  $G_{D^*N} \cdot g_{R,D^*N}$   \\
\hline
$\Lambda_c(2595)$  & $13.88 - 1.06i$ & $26.51 + 2.1i$   \\
$\Lambda_c(2625)$ & $0$  & $29.10$
\end{tabular}
\caption{The values of $G_{DN}\cdot g_{R,DN}$ and $G_{D^*N} \cdot g_{R,D^*N}$ from Ref.~\cite{uchino}. The signs of $G_{D^*N} \cdot g_{R,D^*N}$
are changed with respect to Ref.~\cite{uchino} as discussed in the text.}
\label{tab:Gcoupling}
\end{table}

By looking at Eq.~(\ref{eq:Gamma_R}), we can immediately write the ratio of $\Gamma$ for $\Lambda_c(2595)$ and $\Lambda_c(2625)$ production,
\begin{equation}\label{eq:ratio}
  \frac{\Gamma [\Lambda_b \to \pi^- \Lambda_c(2595)]}{\Gamma [\Lambda_b \to \pi^- \Lambda_c(2625)]}=\frac{M_{\Lambda_c(2595)}}{M_{\Lambda_c(2625)}}
  \frac{p_{\pi 1}}{p_{\pi 2}}\frac{\left[ \overline{\sum}\sum |t_R|^2 \right]_1}{\left[ \overline{\sum}\sum |t_R|^2 \right]_2},
\end{equation}
where $p_{\pi 1}$ and $p_{\pi 2}$ are the pion momenta for the $\Lambda_c(2595)$ and $\Lambda_c(2625)$ production respectively, given by Eq.~(\ref{eq:ppi}) and
$\left[ \overline{\sum}\sum |t_R|^2 \right]_{1,2}$ are given by Eqs.~(\ref{eq:t_R3}), (\ref{eq:t_R4}) respectively.
Using the numerical values of Table \ref{tab:Gcoupling}, we find
\begin{equation}\label{eq:ratio2}
  \frac{\Gamma [\Lambda_b \to \pi^- \Lambda_c(2595)]}{\Gamma [\Lambda_b \to \pi^- \Lambda_c(2625)]}=0.76~.
\end{equation}
Experimentally we have \cite{PDG}
\begin{eqnarray}\label{eq:BR}
  &&BR[\Lambda_b \to \pi^- \Lambda_c(2595),~ \Lambda_c(2595) \to \Lambda_c \pi^+ \pi^-] \nonumber \\
  &=& (3.2 \pm 1.4)\times 10^{-4}, \\
 &&BR[\Lambda_b \to \pi^- \Lambda_c(2625),~ \Lambda_c(2625) \to \Lambda_c \pi^+ \pi^-] \nonumber \\
 &=& (3.1 \pm 1.2)\times 10^{-4}.
\end{eqnarray}
Since the $BR$ for $\Lambda^*_c \to \Lambda_c \pi^+ \pi^-$ is 67\% for both resonances, the ratio of partial decay widths for the
$\Lambda_c(2595)$ to $\Lambda_c(2625)$ summing in quadrature the relative errors is given by
\begin{equation}\label{eq:ratio_Exp}
 \left. \frac{\Gamma [\Lambda_b \to \pi^- \Lambda_c(2595)]}{\Gamma [\Lambda_b \to \pi^- \Lambda_c(2625)]}\right|_{\rm Exp.}=1.03 \pm 0.60.
\end{equation}
The value that we get in Eq.~(\ref{eq:ratio2}) is compatible within errors.

We should call the attention to the fact that the $DN$ and $D^*N$ contributions are about the same for the $\Lambda_c(2595)$ case and sum constructively.
Should the sign be opposite then there would be a near cancellation of the rate for the case of $\Lambda_c(2595)$ and there would have been massive disagreement with experiment. This point is worth mentioning because in Ref.~\cite{uchino} the signs for the $D^*N$ couplings are opposite to those in Table \ref{tab:Gcoupling}. The reason for the change of sign here is that in Ref.~\cite{uchino} a full box diagram with $\pi$ exchange on each side was evaluated. This provided the value of $V^2_{\rm eff}$ to be used in coupled channels of $DN$ and $D^*N$ and, since the sign did not matter for the spectra discussed in Ref.~\cite{uchino} the positive sign of $V_{\rm eff}$ was chosen by default. The sign here is crucial and hence, taking the negative sign for $V_{\rm eff}$, as it corresponds to $\pi$ exchange, is the correct choice. The signs then also agree with those obtained in Ref.~\cite{romanets} just using symmetries.

We can now make prediction for the reactions $\Lambda_b \to D^-_s \Lambda_c(2595)$ and $\Lambda_b \to D^-_s \Lambda_c(2625)$. The reactions are analogous.
It suffices to substitute the $W^- \bar ud$ vertex in Fig. \ref{fig:FeynmanDiag1} by $W^- \bar cs$, which is also Cabibbo favoured and goes with $\cos \theta_C$ as in the $\pi^-$ case. Thus, the formulae for the widths are identical changing the kinematics to account for the larger $D^-_s$ mass. Yet, given the large mass of the $\Lambda_b$ and the available phase space, the momenta of the pseudoscalar mesons, and particularly the energies are not too different to those in the $\pi$ case.

We can construct the ratio of Eq.~(\ref{eq:ratio}) for the $D^-_s$ case and we find
\begin{equation}\label{eq:ratio_Ds}
  \frac{\Gamma [\Lambda_b \to D^-_s \Lambda_c(2595)]}{\Gamma [\Lambda_b \to D^-_s \Lambda_c(2625)]}=0.54~.
\end{equation}
This is a good prediction that relies upon the $ME(q)$ being about the same for the decay into $\Lambda_c(2595)$ and $\Lambda_c(2625)$.

Assuming that $ME(q)$ is the same for $\Lambda_b \to D^-_s \Lambda_c(2595)$ and $\Lambda_b \to \pi^- \Lambda_c(2595)$, we can make another prediction but with larger error. Actually $q_{D_s} = 1630 ~{\rm MeV}/c$ and $q_{\pi} = 2208 ~{\rm MeV}/c$, so $ME(q)$ is not necessarily equal, but we can provide some estimate of the rate assuming the same value for $ME(q)$. Then we have
\begin{eqnarray}\label{eq:GamDspi}
&&\Gamma[\Lambda_b \to D^-_s \Lambda_c(2595)]  \nonumber \\
&=& \frac{q_{D_s} (q^2_{D_s}+w^2_{D_s})}{q_{\pi} (q^2_{\pi}+w^2_{\pi})} \;\Gamma[\Lambda_b \to \pi^- \Lambda_c(2595)], \\
&&\Gamma[\Lambda_b \to D^-_s \Lambda_c(2625)] \nonumber \\
&=& \frac{q_{D_s}\cdot 2w^2_{D_s}}{q_{\pi} \cdot 2w^2_{\pi}} \;\Gamma[\Lambda_b \to \pi^- \Lambda_c(2625)],
\end{eqnarray}
and we obtain
\begin{eqnarray}%\label{eq:BR_Ds2}
  BR[\Lambda_b \to D^-_s \Lambda_c(2595)] &\sim& (2.22\pm 0.97)\times 10^{-4},~~~~ \label{eq:BR_Ds2-1}\\
  BR[\Lambda_b \to D^-_s \Lambda_c(2625)] &\sim& (3.03\pm 1.70)\times 10^{-4},~~~~\label{eq:BR_Ds2-2}
\end{eqnarray}
where in Eqs.~(\ref{eq:BR_Ds2-1}),(\ref{eq:BR_Ds2-2}) we have taken the experimental rates for $\Lambda_b \to \pi^- \Lambda_c(2595)$ and $\Lambda_b \to \pi^- \Lambda_c(2625)$ with their errors. Since the momenta of the $D^-_s$ is smaller than the one of $\pi^-$ and $ME(q)$ should decrease with $q$, we could expect the values of Eqs.~(\ref{eq:BR_Ds2-1}), (\ref{eq:BR_Ds2-2}) to be lower limits, but given the large errors, the order of magnitude of these numbers should be relatively accurate.

\section{Conclusions}
We have studied the $\Lambda_b \to \pi^- \Lambda_c(2595)$ and $\Lambda_b \to \pi^- \Lambda_c(2625)$ from the perspective that the $\Lambda_c(2595)$ and $\Lambda_c(2625)$ are dynamically generated resonances from the interaction of $DN, D^*N$ with coupled channels. We have developed a formalism to relate the
$\Lambda_b \to \pi^- DN$ and $\Lambda_b \to \pi^- D^*N$ decays. For this we make a detailed model of the $\bar qq$ hadronization, using the $^3P_0$ picture for the creation of $\bar qq$ with the quantum numbers of the vacuum. Racah algebra is used to relate these couplings and final easy expressions are obtained. This, together with the couplings of the resonances $\Lambda_c(2595)$ and $\Lambda_c(2625)$ to $DN$ and $D^*N$ obtained before in a full coupled channel approach, including $DN$ and $D^*N$, allows us to obtain the decay rates up to an unknown global factor related to the matrix element of the radial wave functions of the $b$ and $c$ quarks. The ratio of rates is then a prediction of the theory and is in good agreement with experiment within experimental uncertainties. We could also obtain the ratio of rates for $\Lambda_b \to D^-_s \Lambda_c(2595)$ and $\Lambda_b \to D^-_s \Lambda_c(2625)$, which are not measured so far. We also made estimates of the branching fractions for these two latter decays, not only their ratio.

One of the important findings of the work was the relevance of the $D^*N$ component in the $\Lambda_c(2595)$, which was overlooked in early works studying these resonances. We found that the $D^*N$ had a strength similar to that of the $DN$ component and was essential to have good agreement with experiment. Also, the relative sign of the coupling of the $\Lambda_c(2595)$ to $DN$ and $D^*N$ was of crucial importance. An opposite sign to the one that we obtain leads to large cancellations of the $\Lambda_b \to \pi^- \Lambda_c(2595)$, such that there would be an absolute disagreement with the data.

The mixture of pseudoscalar-baryon and vector-baryon states in coupled channels is catching up \cite{garzon,vectorreview,uchino,uchino2,kanchan,kanchan2} and,
as done in the present work, it would be interesting to find similar reactions that evidence the relevance of this mixing.

\section*{Acknowledgments}
We thank Juan Nieves for a critical reading of the paper and valuable comments.
This work is partly supported by the
National Natural Science Foundation of China under Grants
No.~11565007 and No.~11547307.
This work is also partly supported by the Spanish Ministerio
de Economia y Competitividad and European FEDER funds
under the contract number FIS2011-28853-C02-01, FIS2011-
28853-C02-02, FIS2014-57026-REDT, FIS2014-51948-C2-
1-P, and FIS2014-51948-C2-2-P, and the Generalitat Valenciana
in the program Prometeo II-2014/068.

%\clearpage

\bibliographystyle{plain}

\end{document}